\documentclass[12pt]{article}

\usepackage{amssymb}
\usepackage{color,graphicx}
\usepackage{amsmath}
\usepackage{hyperref}

\begin{document}

\renewcommand{\abstractname}{\hfill}

\newpage
\pagenumbering{arabic}
\begin{center}
\bf
A. A. Grib${}^*$,\footnote{${}^*$Theoretical Physics and Astronomy Department
of the Herzen  University
and
A.~Friedmann Laboratory for Theoretical Physics,
Saint Petersburg, Russia,

E-mail:\, andrei\_grib@mail.ru}
Yu. V. Pavlov${}^{**}${}\footnote{${}^{**}$Institute of Problems in
Mechanical Engineering of Russian Academy of Sciences
and
A.~Friedmann Laboratory for Theoretical Physics, Saint Petersburg, Russia,

E-mail:\, yuri.pavlov@mail.ru}
\end{center}

\begin{center}
\Large \bf High energy physics in the vicinity\\ of rotating black holes
\end{center}

\begin{abstract}
    The aim of the paper is the consideration of particle collisions in
the vicinity of the horizon of rotating black holes.
    Existence of geodesics for massive and massless particles arising from
the region inside of the gravitational radius in ergosphere leads to
the different possibilities to get very high energy in
the centre mass frame of two particles.
    The classification of all such geodesics on the basis of the proved
theorem for extremal spherical orbits is given.
    Case of the unlimited growth of energy for the situation when one of
the  particles (the critical one) is moving along the ``white hole'' geodesic
with the close to the upper limit angular momentum while the other particle
is on the usual geodesic and the case of the unlimited negative
angular momentum of the first particle are considered.
\end{abstract}


{\centering \section{{\large Introduction}}}

    In the paper of Ba\~{n}ados, Silk and West~\cite{BanadosSilkWest09}
the existence of the special resonance in the vicinity of the horizon of
the rotating black hole with extremal angular momentum was shown leading
to the unlimited growth of the energy in the centre of mass frame.
    The physical reason of the appearance of such a resonance is simple.
        One of the particles called the critical one is strongly ``rotated''
in the ergosphere of the rotating black hole so that its radial component
is close to zero.
    The other noncritical particle has the radial component close to the
velocity of light.
    So the relative velocity of two particles also is close to the velocity
of light and unlimited growing Lorentz factor leads to the unlimited growth
of the energy in the centre of mass frame.

    In our papers~\cite{GribPavlov2010}--\cite{GribPavlov2011b}
the similar effect was considered for the nonextremal Kerr's black holes due
to the opinion that extremal black holes are absent in Nature due to
gravitational radiation of such objects~\cite{Thorne74}.
    It was shown that critical particles can arise due to multiple
collisions when noncritical  particle arriving to the horizon from
the external space after collision with another noncritical particle
acquires the critical  angular momentum after the intermediate collision.

    At last in our papers~\cite{GribPavlov2012,GribPavlov2013}
it was shown  that growing without limit high energies in the centre of mass
frame in collisions of two particles can occur at any point of the ergosphere
if one of the particles has the large negative angular momentum.
    However there was a problem to get such a large momentum.

    A new phenomenon studied by us in~\cite{GribPavlov2015} is the role of
white hole geodesics in collisions of particles in the ergosphere of
the black hole.
    As it is known from the text books~\cite{LL_II} the geodesic completeness
leads necessarily to the existence together with usual geodesics arriving
from the external space the geodesics coming from the inside of
the gravitational radius called by us the ``white hole geodesics''.
     It occurs that all geodesics with negative energies relative to
the infinity~\cite{GribPavlovVert} playing important role in
the Penrose process~\cite{Penrose69} of getting the energy from
the rotating black hole belong to this class.

    The present paper is the continuation of~\cite{GribPavlov2015}.
    Differently from~\cite{GribPavlov2015} where only the equatorial geodesics
were considered here we prove the theorem leading to the classification of
all such lines in the general case.
    Massless particles on white hole geodesics are also considered.
    The unlimited growth of the energy of collision in the centre of
mass frame is considered for the case when one of the colliding particles
is moving along the white hole geodesic.
    It is shown that both effects considered in our previous papers can
occur without additional hypotheses on getting the critical momentum
or large in absolute value angular momentum.

   The system of units $G=c=1$ is used.

\vspace{4mm}
{\centering \section{{\large The white hole geodesics in Kerr's metric}}
\label{sec2}}

    The Kerr's metric~\cite{Kerr63} in Boyer-Lindquist
coordinates~\cite{BoyerLindquist67} has the form:
    \begin{equation}
d s^2 = d t^2 -
\frac{2 M r}{\rho^2} \, ( d t - a \sin^2 \! \theta\, d \varphi )^2
\label{Kerr}
- \rho^2 \left( \frac{d r^2}{\Delta} + d \theta^2 \right)
- (r^2 + a^2) \sin^2 \! \theta\, d \varphi^2,
\end{equation}
    where
    \begin{equation} \label{Delta}
\rho^2 = r^2 + a^2 \cos^2 \! \theta, \ \ \ \ \
\Delta = r^2 - 2 M r + a^2,
\end{equation}
    $M$ is the black hole mass, $ aM $ its angular momentum.
        Let ${0 \le a \le M }$.
    The event horizon of the Kerr's black hole corresponds to the coordinate
    \begin{equation}
r = r_H \equiv M + \sqrt{M^2 - a^2} .
\label{Hor}
\end{equation}
    The static limit surface is defined by the value
    \begin{equation}
r = r_1 \equiv M + \sqrt{M^2 - a^2 \cos^2 \theta} .
\label{Lst}
\end{equation}
    The region of space-time between the static limit and the event horizon
is called ergosphere.

    The geodesic equations in Kerr's metric~(\ref{Kerr}) have the
form~(see~\cite{Chandrasekhar}, Sec.~62 or~\cite{NovikovFrolov}, Sec.~3.4.1)
    \begin{equation} \label{geodKerr1}
\rho^2 \frac{d t}{d \lambda } = -a \left( a E \sin^2 \! \theta - J \right)
+ \frac{r^2 + a^2}{\Delta} P, \ \
\rho^2 \frac{d \varphi}{d \lambda } =
- a E + \frac{J}{\sin^2 \! \theta} + \frac{a P}{\Delta} ,
\end{equation}
    \begin{equation} \label{geodKerr3}
\rho^2 \frac{d r}{d \lambda} = \sigma_r \sqrt{R}, \ \ \ \ \
\rho^2 \frac{d \theta}{d \lambda} =\sigma_\theta \sqrt{\Theta},
\end{equation}
    \begin{equation} \label{geodR}
R = P^2 - \Delta [ m^2 r^2 + (J- a E)^2 + Q], \ \ \
P = \left( r^2 + a^2 \right) E - a J,
\end{equation}
    \begin{equation} \label{geodTh}
\Theta = Q - \cos^2 \! \theta \left[ a^2 ( m^2 - E^2) +
\frac{J^2}{\sin^2 \! \theta} \right].
\end{equation}
    Here $E={\rm const}$ is the energy (relative to the infinity) of
the moving particle,
$J$ is the conserved projection of its angular momentum on the axis
of rotation of the black hole,
$m$ is the rest mass of the moving particle.
    For a particle with $m \ne 0$ the parameter $\lambda = \tau /m$,
where $\tau$ is the proper time.
    $Q$ is the Carter constant.
    For the movement in equatorial plane $Q=0$.
    The constants $\sigma_r, \sigma_\theta = \pm 1$ define the direction
of movement in coordinates $r, \theta$.

    As it follows from~(\ref{geodKerr3}) the parameters characterizing
any geodesic must satisfy the conditions
    \begin{equation} \label{ThB0}
R \ge 0, \ \ \ \ \ \Theta \ge 0 .
\end{equation}
    The condition of movement of the particle ``forward'' in time
must be valid, so
    \begin{equation} \label{ThB0t}
d t / d \lambda \ge 0 .
\end{equation}
    The conditions~(\ref{ThB0}), (\ref{ThB0t}) lead to the following
inequalities for possible values of the energy $ E $ and
the angular momentum projection $J$ for the test particle at the point
with coordinates $ (r, \theta) $    for the fixed
value of~$\Theta \ge 0$~\cite{GribPavlov2013}:

    Outside the ergosphere $ r^2 -2 r M +a^2 \cos^2 \! \theta >0 $ one has
    \begin{equation} \label{EvErg}
E \ge \frac{1}{\rho^2} \sqrt{(m^2 \rho^2 + \Theta)
(r^2 -2 r M +a^2 \cos^2 \! \theta)},
\ \  \ J \in \left[ J_{-} (r,\theta), \ J_{+} (r,\theta) \right],
\end{equation}
    \begin{eqnarray}
J_{\pm} (r,\theta) = \frac{\sin \theta}{r^2 -2 r M +a^2 \cos^2 \! \theta}
\biggl[ - 2 r M a E \sin \theta
\nonumber \\
\pm \, \sqrt{ \Delta \left( \rho^4 E^2 - (m^2 \rho^2 + \Theta)
(r^2 -2 r M +a^2 \cos^2 \! \theta) \right)} \biggr] .
\label{Jpm}
\end{eqnarray}

    Inside the ergosphere  $ r_H < r < r_1(\theta) $ one has
    \begin{equation} \label{lHmdd}
r^2 -2 r M +a^2 \cos^2 \! \theta <0 , \ \ \ \
J \le J_{+}(r,\theta) \le J_{-}(r,\theta),
\end{equation}
    and the value of the particle energy can be any, as positive as negative.
        For~$ r $ close to the horizon $ r_H $ from~(\ref{Jpm}), (\ref{lHmdd})
(for $\theta \ne 0, \pi$) one obtains
    \begin{equation} \label{JgEH}
J \le J_H = \frac{ 2 M r_H E}{a} .
\end{equation}
    So $J_H$ is the upper frontier for the angular momentum projection of
the particle with the energy~$E$ close to the event horizon of the black hole.

    Earlier in paper~\cite{GribPavlovVert} it was shown that geodesics with
negative relative to the infinity energy in ergosphere originate and terminate
at $r=r_H$.
    And what is about the geodesics with positive energy also originating and
terminating  at $r=r_H$?
    Let us consider this problem?

    Introduce the effective potential for the radial movement
    \begin{equation} \label{Leff}
V_{\rm eff} = - \frac{R}{2 \rho^4}.
\end{equation}
    Then due to~(\ref{geodKerr3}),
    \begin{equation} \label{LeffUR2}
\frac{1}{2} \left( \frac{d r}{d \lambda} \right)^{\!2} + V_{\rm eff}=0 ,
\ \ \ \frac{d^2 r}{d \lambda^2} = - \frac{\partial V_{\rm eff}}{\partial r} .
\end{equation}
    To find the answer on the posited question it is sufficient to find
the regions of energy values of particles such that
    \begin{equation} \label{LeffdVefg0}
r > r_H, \ \ \ V_{\rm eff}(r)=0 \ \ \Rightarrow
\ \ \frac{\partial V_{\rm eff}}{\partial r} > 0 .
\end{equation}
    Really the condition~(\ref{LeffdVefg0}) means that in the upper in $r$
point of the trajectory the acceleration $ d^2 r/ d \lambda^2 $
due to~(\ref{LeffUR2}) is directed in the direction of decrease of the radial
coordinate i.e. to the event horizon.
    To find the boundary values of the parameters for such ``white hole''
geodesics consider the system of equations
    \begin{equation} \label{krug}
V_{\rm eff}(r)=0, \ \ \ \
V'_r \equiv \frac{\partial V_{\rm eff}}{\partial r} = 0 .
\end{equation}
    This system defines the orbits with constant value of the radial
coordinate --- the so called spherical orbits~\cite{Dymnikova86}.

    For lightlike geodesics ($m=0$) the system of equations~(\ref{krug})
leads to expressions
    \begin{equation} \label{cuclm0}
\frac{J}{E} = \frac{r^2 (3M-r) - a^2 (r+M)}{a(r-M)} ,
\ \ \ \
\frac{a^2 Q}{E^2} = \frac{r^3 \left( 4 M a^2 - r (r-3M)^2 \right)}{( r-M)^2} .
\end{equation}
    For the nonrotating black hole ($a=0$) the spherical photon orbits are
circular with values
    \begin{equation} \label{cum0Sch}
J = \pm 3 \sqrt{3} M E, \ \ \ \ r = 3 M .
\end{equation}
    In the region $ 2M < r <3M$ inside the radius of the circular photon
orbit of the Schwarzschild metric all lightlike geodesics
with $ |J| / M E > 3 \sqrt{3} $ originate and terminate at the gravitational
radius $r_H = 2 M $.

\begin{figure}[ht]
\centering
\includegraphics[width=65mm]{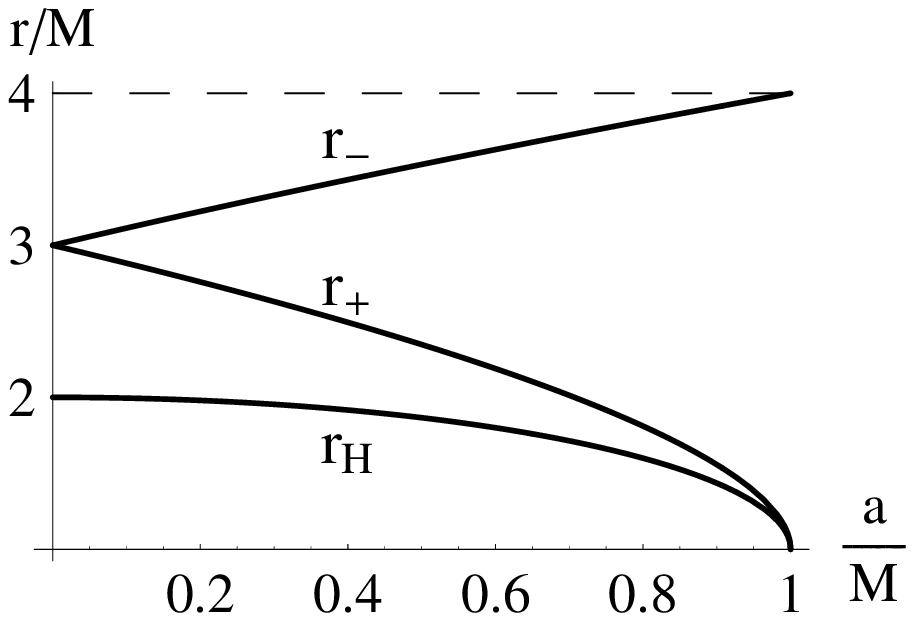} \
\includegraphics[width=65mm]{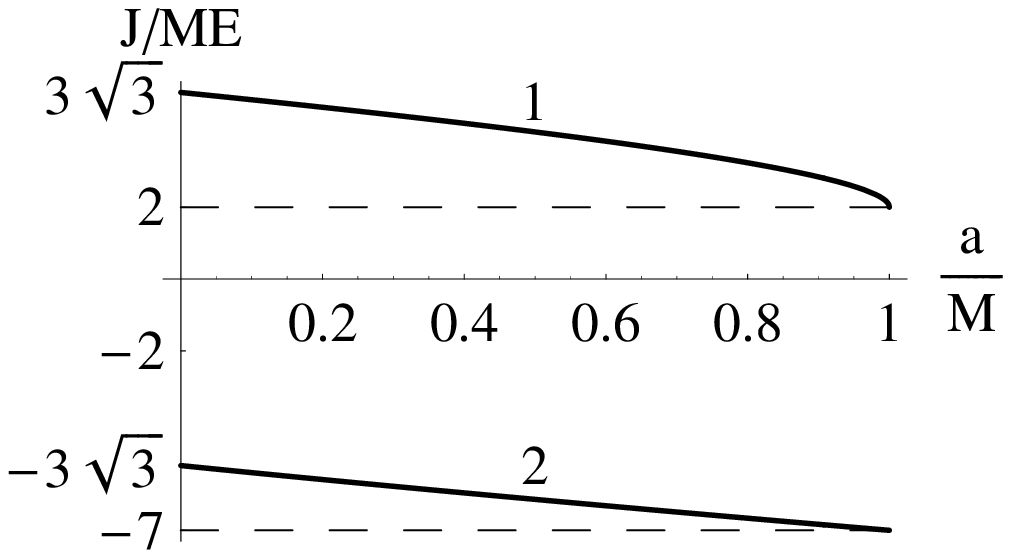} \
\caption{The radii (to the left) and angular momenta (to the right)
of circular photon orbits in Kerr's metric.}
\label{FigGr1}
\end{figure}

    For the rotating Kerr's black hole the radii of equatorial photonic
circular orbits $Q=0$ can be found from~(\ref{cuclm0}) as roots of
the equation $r (r-3M)^2 - 4 M a^2 =0 $ and out the event horizon they are
equal to~\cite{BardeenPressTeukolsky72}
    \begin{equation} \label{cum0K1}
r_{\pm} = 2 M \left[ 1 + \cos \left( \frac{2}{3} \arccos \frac{\mp a}{M}.
\right) \right]
\end{equation}
    The corresponding values of $ r_{\pm} $ and $J / M E $
are represented on Fig.~\ref{FigGr1}.
    One can see that the obtained values of $J / M E $ coincide with
the limiting values of $ J_\pm (r_\pm, \pi/2) / M E $ calculated using
formulas~(\ref{Jpm}).
    That is why if in the region $ r_H < r < r_+ $ the values of $J / M E $
are larger than the values given by the curve~({1}) on Fig.~\ref{FigGr1}
to the right then such geodesics originate and terminate on the gravitational
radius~$r_H$.
    If in the region $ r_H < r < r_- $ the values of $J / M E $
are less than the value given by the curve~({2}) on Fig.~\ref{FigGr1}
to the right then such geodesics originate and terminate on
the gravitational radius~$r_H$.
    Note that due to~(\ref{JgEH}) close to the horizon
one has $J / M E < 2 r_H/ a$.

\vspace{4mm}
{\centering \section{\large Properties of spherical orbits in Kerr's metric}
\label{secSpOrb}}

    As it was shown earlier the problem of boundary values of the parameters
outside of the event horizon for geodesics originating and terminating
on~$r=r_H$ is identical to the search of boundary values of parameters
for spherical orbits.
    In case of massive particles $m \ne 0 $ the minimal value of the energy
for the spherical orbits limits the maximal value of the energy for such
white hole geodesics.

    The system of equations~(\ref{krug}) defining the spherical orbits
can be written as
    \begin{equation} \label{LeffCucl}
R(r,m,E,J,Q)=0, \ \ \ \ \frac{\partial R (r,m,E,J,Q)}{\partial r} = 0 .
\end{equation}
    So there is no dependence of the effective potential on
the coordinate~$\theta$.

    For fixed $m \ne 0 $ the conditions~(\ref{LeffCucl}) define
(probably non uniquely) the values of the energy and of the projection of
the angular momentum as functions of the radial coordinate $r$ and
the Carter constant~$Q$:
    \begin{equation} \label{les}
E = E (r, Q), \ \ \ \ J = J (r, Q) .
\end{equation}
    Differentiating the equalities~(\ref{krug}) with respect to $r$
due to~(\ref{les}) one obtains
    \begin{equation} \label{les2}
\frac{\partial V }{\partial r} +
\frac{\partial V}{\partial E } \frac{\partial E}{\partial r} +
\frac{\partial V}{\partial J } \frac{\partial J}{\partial r} =0 ,
\end{equation}
    \begin{equation} \label{les3}
\frac{\partial^2 V }{\partial r^2} +
\frac{\partial V^\prime_r}{\partial E } \frac{\partial E}{\partial r} +
\frac{\partial V^\prime_r}{\partial J } \frac{\partial J}{\partial r} = 0 .
\end{equation}

    If one looks for the extremal on angular momentum spherical orbit for
fixed value of  $Q$, i.e. $\partial J / \partial r =0$
then from~(\ref{les2}) one obtains
    \begin{equation} \label{les4a}
V =0 , \ \ \ V^\prime_r = 0, \ \ \
\frac{\partial J (r,Q) }{\partial r} = 0 \ \ \Rightarrow \ \
\frac{\partial V}{\partial E } \frac{\partial E}{\partial r} = 0 .
\end{equation}
    One can show that for any test particle in Kerr's metric
    \begin{equation} \label{les4aa}
V =0 , \ \ \ V^\prime_r = 0, \ \ \
\frac{\partial J (r,Q) }{\partial r} = 0 \ \ \Rightarrow \ \
\frac{\partial V}{\partial E } \ne 0 .
\end{equation}
    So the extremal in angular momentum spherical orbit is also extremal
in the energy value: $\partial E / \partial r =0$.

    If one looks for the extremal in energy spherical orbit for fixed
values of  $Q$, i.e. $\partial E / \partial r =0$ then from~(\ref{les2})
one has
    \begin{equation} \label{les4}
V =0 , \ \ \ V^\prime_r = 0, \ \ \
\frac{\partial E (r,Q) }{\partial r} = 0 \ \ \Rightarrow \ \
\frac{ \partial V }{\partial J }
\frac{ \partial J }{\partial r} = 0 .
\end{equation}
    Elementary but long calculations give
    \begin{eqnarray}
V =0 , \ \ \ V^\prime_r = 0, \ \ \
\frac{\partial E (r,Q) }{\partial r} = 0 , \ \ \
\frac{\partial V}{\partial J } = 0  \ \ \Rightarrow \hspace{17mm}
\nonumber \\
a\ne 0, \ m \ne 0, \ \ r = 6 M, \ \ E = \frac{2 \sqrt{2}}{3} m, \ \
J = - \frac{\sqrt{2}}{3} m a, \ \ Q = 12 m^2 M^2 .
\label{les4b}
\end{eqnarray}
    If one is not interested in this exception then the extremal in
the energy spherical orbit is also extremal in the angular momentum
projection: ${\partial J / \partial r =0}$.

    From~(\ref{les3}) one obtains
    \begin{equation} \label{les5}
V =0 , \ \ \ V^\prime_r = 0, \ \ \
\frac{\partial E }{\partial r} = 0, \ \
\frac{\partial J }{\partial r} = 0 \ \ \Rightarrow \ \
\frac{\partial^2 V }{\partial r^2} = 0 .
\end{equation}
    If for fixed parameters of the geodesic $(m,E,J,Q)$ and for small deviation
of $r$ corresponding to the spherical orbit the force $d^2 r/ d \lambda^2$
appears tending to lead the value of $r$ to the initial value,
then such orbits are called steady.
    From~(\ref{LeffUR2}) and (\ref{krug}) it follows that the spherical orbit
is steady if
    \begin{equation} \label{ust}
V_{\rm eff}(r)=0, \ \ \ \ \frac{\partial V_{\rm eff}}{\partial r} = 0,
\ \ \ \ \frac{\partial^2 V_{\rm eff}}{\partial r^2} > 0 .
\end{equation}
    So due  to~(\ref{les5}) all extremal in energy
(with exception~(\ref{les4b})) spherical orbits are also limiting stable
spherical orbits.

    One can formulate this as the {\bf theorem on extremal orbits}.
    {\it The spherical orbit in the field of the Kerr's black hole extremal
in the value of the angular momentum projection (in energy with the exception
of the case~(\ref{les4b})) for the fixed value of the Carter constant
is also extremal in the value of the energy (the angular momentum projection)
and it is the stable circular orbit.}

    For $Q=0$ and $\theta=\pi/2$ one obtains the evident {\bf consequence}.
    {\it The circular orbit in Kerr's metric, extremal in the value of
the energy (angular momentum) is also extremal in the value of the angular
momentum (energy) and is the stable circular orbit.}

    For the Newton gravitational law the equations~(\ref{les})
in usual units have the form
    \begin{equation} \label{LNyu2}
E = m c^2 - G \frac{m M}{2 r} , \ \ \
J = \pm m \sqrt{G M r} .
\end{equation}
    All circular orbits are stable.
    Extremal values in energy and angular momentum occur only on
the boundaries for $r \to \infty $ or $r \to 0 $.

    The case of the Schwarzschild metric was considered by us
in~\cite{GribPavlov2015} where it was shown that for particle
energies $E < m 2 \sqrt{2}/3$, the geodesics out of the  event horizon of
the black hole originate and terminate on the gravitational radius $r = 2M$.

    Solving the system of equations~(\ref{krug}) for Kerr's metric after
long algebraic transformations one obtains
    \begin{equation} \label{NFore}
\varepsilon \equiv \! \frac{E}{m} \!=\! \frac{x^3 (x-2) + A H_{\pm}}
{x^2 \sqrt{x^3(x-3) + 2 A H_{\pm} } } , \ \
l \equiv \! \frac{J}{m M} \!=\! \frac{(x^2 + A^2) H_{\pm} - 2 A x^3 }
{x^2 \sqrt{x^3(x-3) + 2 A H_{\pm} } },
\end{equation}
    where
    \begin{equation} \label{xAF}
x =\! \frac{r}{M} , \ \ A =\! \frac{a}{M} ,\ \ q =\! \frac{Q}{m^2 M^2} , \ \
H_{\pm} = \pm \sqrt{x^5 + A^2 q^2 + q x^3 (3-x)} - A q .
\end{equation}
    For $Q=0$  formulas~(\ref{NFore}) give known expressions for equatorial
orbits of the Kerr's metric~\cite{BardeenPressTeukolsky72}.
    The results~\cite{GribPavlov2015} obtained for this case  from
formulas~(\ref{NFore}) follow from the proved by us theorem and known
parameters of limiting circular stable orbits.
    For example from the value of the energy $E = m/\sqrt{3}$ of the stable
limiting orbit of the black hole with $a=M$ it follows that all equatorial
geodesics with the energy $E < m/\sqrt{3}$ out of the horizon originate and
terminate on $r =r_H $.

    Differentiating~(\ref{NFore}) one obtains
    \begin{equation} \label{prEq}
\frac{\partial \varepsilon}{\partial q} =
\frac{\pm A (x^3 - A H_\pm)}{2 x^2 \sqrt{x^3(x-3) + 2 A H_{\pm} }
\sqrt{x^5 + A^2 q^2 + q x^3 (3-x)} } .
\end{equation}
    For arbitrary geodesics with $E < m$ the Carter's constant $Q\ge 0$
due to~(\ref{geodKerr3}), (\ref{geodTh}).
    In this case positive values of~$J$ for spherical orbits are possible
only for the choice of $H_+$ in~(\ref{NFore}).
    Then from~(\ref{prEq}) one obtains
$ \partial \varepsilon / \partial q > 0 $ for $r>r_H$ and so for the extremal
spherical orbits ($ \partial \varepsilon / \partial x = 0 $)
    \begin{equation} \label{dprEq}
\frac{d \varepsilon}{d q} =
\frac{\partial \varepsilon}{\partial x} \frac{d \varepsilon}{\partial q} +
\frac{\partial \varepsilon}{\partial q} =
\frac{\partial \varepsilon}{\partial q} > 0 .
\end{equation}
    On the boundaries of the region of existence of spherical orbits
with $H_+>0$ for $r >r_H$ one can see that the energy cannot be larger than
the corresponding values for equatorial circular orbits.
    That is why the limiting energy values of energies for massive particles
found for equatorial geodesics in~\cite{GribPavlov2015} are the same
for the nonequatorial movement with $J>0$, particularly all geodesics with
the energy $E < m/\sqrt{3}$ and $J>0$ out of the horizon of the Kerr's black
hole with $a=M$ originate and terminate on $r =r_H $.

\vspace{4mm}
{\centering \section{\large The energy of particles collision in
the centre of mass frame}
\label{secX}}

    One can find the energy $E_{\rm c.m.}$ of two colliding particles with
masses~$m_1$ and $m_2$ in the centre of mass frame by taking the square
of the formula
    \begin{equation} \label{SCM}
\left( E_{\rm c.m.}, 0\,,0\,,0\, \right) = p^{\,i}_{(1)} + p^{\,i}_{(2)},
\end{equation}
    where $p^{\,i}_{(n)}$ are 4-momenta of particles $(n=1,2)$.
    Taking the square of~(\ref{SCM})
and using~(\ref{geodKerr1}), (\ref{geodKerr3}) and
$p^{\,i}_{(n)} p_{(n)i}= m_n^2$ one obtains~\cite{HaradaKimura11}:
    \begin{eqnarray}
E_{\rm c.m.}^{\,2} &=& m_1^2 + m_2^2 +
\frac{2}{\rho^2} \biggl[ \, \frac{P_1 P_2 -
\sigma_{1 r} \sqrt{R_1} \, \sigma_{2 r} \sqrt{R_2}}{\Delta}
\nonumber \\
&&-\, \frac{ (J_1 - a  E_1 \sin^2 \! \theta) (J_2 - a  E_2 \sin^2 \! \theta)}
{\sin^2 \! \theta}
- \sigma_{1 \theta} \sqrt{\Theta_1} \,
\sigma_{2 \theta} \sqrt{\Theta_2} \, \biggr].
\label{KerrL1L2}
\end{eqnarray}
    Taking the limit for $r \to r_H$, one obtains for falling
$\sigma_{1 r} = \sigma_{2 r}=1$ particles in equatorial ($\theta= \pi/2$)
plane~\cite{GribPavlovPiattella2012}:
    \begin{eqnarray}
E_{\rm c.m.}^{\,2} (r \to r_H) = \frac{ (J_{1H} J_2 - J_{2H} J_1)^2}
{4 M^2 (J_{1H} - J_1) (J_{2H} - J_2)}
\nonumber \\
+\, m_1^2 \left[1+ \frac{J_{2H} - J_2}{J_{1H} - J_1}\right] +
m_2^2 \left[ 1+ \frac{J_{1H} - J_1}{J_{2H} - J_2}\right],
\label{KerrLime1e2e2f}
\end{eqnarray}
    So the energy of the collision in the centre of mass frame is increasing
without limit if $r \to r_H$ and the angular momentum for the fixed relative
to infinity energy of one of the particles (the critical) tend to the limiting
value close to the event horizon $J_H$ (see~(\ref{JgEH})).
    However, excluding the case of the extremal rotating black hole $a=M$
particles with the angular momentum $J_H$ cannot achieve the horizon if they
fall from the infinity.
    The effective potential energy creates the barrier so high energies of
collisions in the centre of mass frame can be achieved only in decays or
multiple collisions~\cite{GribPavlov2010,GribPavlov2011}, when one has the
increasing of the angular momentum of one of the falling particles $J \to J_H$.
    Other possibility of the unlimited growth of energy of collisions close
to the event horizon is  the collision with the particles moving in the
opposite direction moving along the white hole geodesics~\cite{GribPavlov2015}.
    However as one can see from the Penrose diagram for eternal black holes
for the Schwarzschild case for $r\to r_H$ such collisions could occur in
the infinite past.

    There is another possibility of collisions with unlimited energy.
    From ~(\ref{KerrL1L2}) in the ergosphere in the limit $J_2 \to -\infty$
one obtains
    \begin{eqnarray}
E_{\rm c.m.}^{\,2} \approx J_2 \frac{r^2\! -2 r M \!+a^2 \cos^2 \! \theta}
{\rho^2 \Delta \sin^2 \! \theta}
\left( \sigma_{1r} \sqrt{J_{1 +}\!\!- J_1} -
\sigma_{2r} \sqrt{J_{1 -}\!\!- J_1} \right)^2 \!\to + \infty.
\label{KerrJBner}
\end{eqnarray}
    The energy of collisions is increasing without limit (for fixed energy
of particles relative to infinity) when the angular momentum of one of
the particles in the ergosphere tends to $ -\infty$.
    Due to the fact that particle with $J_2 \to -\infty$ cannot arrive
to ergosphere there is a necessity of multiple
collisions~\cite{GribPavlov2012,GribPavlov2013}
or existence of strong magnetic fields.

    However if the collision of the particle coming from the infinity occurs
with the particle moving along the white hole geodesic for which
due to~(\ref{lHmdd}) there are no limitations from below on the angular
momentum then the energy of collisions in the centre of mass frame can be
large without limit at any point of the ergosphere.

    For white hole geodesics close to the horizon the upper limit of
the permitted angular momentum of particles tend to $J_H$ (see~(\ref{JgEH})).
    The energy of collision~(\ref{KerrLime1e2e2f}) close to the horizon of
the particle falling from the infinity with that moving along the white hole
geodesic with such angular momentum will be also high.
    There is no necessity of multiple collisions.

{\bf Acknowledgement.}\,
The work was supported by RFBR, grant~15-02-06818-a.

\vspace{4mm}

\end{document}